\documentclass[aps,prd,showpacs,amsmath,amssymb]{revtex4}

\usepackage{graphicx}% Include figure files
\usepackage{dcolumn}% Align table columns on decimal point
\usepackage{bm}% bold math

\begin{document}

%\preprint{APS/123-QED}

\title{CPT and lepton number violation in neutrino sector:
Modified mass matrix and oscillation due to gravity}

\author{Monika Sinha}
 \email{monika.sinha@saha.ac.in}
\affiliation{%
Theory Division, Saha Institute of Nuclear Physics \\ 1/AF
Bidhannagar, Kolkata 700 064, India.
}%

\author{Banibrata Mukhopadhyay}
 \email{bm@physics.iisc.ernet.in}
\affiliation{ Department of Physics, Indian Institute of Science
\\
Bangalore 560012, India.
}%

%\date{\today}

\begin{abstract}
We study the consequences of CPT and lepton number violation in
neutrino sector. For CPT violation we take gravity with which
neutrino and antineutrino couple differently. Gravity mixes
neutrino and antineutrino in an unequal ratio to give two mass
eigenstates. Lepton number violation interaction together with
CPT violation gives rise to neutrino-antineutrino oscillation. Subsequently,
we study the neutrino flavor mixing and oscillation under the influence of gravity.
It is found that gravity changes flavor oscillation significantly which
influences the relative abundance of different flavors in present universe.
%The right estimate of neutrino abundance may throw light on the open question of dark energy.
We show that
%although CPT violation signifies different mixing
%ratio and different effective masses for neutrino-antineutrino sector,
the neutrinoless double beta decay rate is modified due to
presence of gravity$-$ the origin of CPT violation, as the mass of the
flavor state is modified.
\end{abstract}


 \maketitle

\section{Introduction}

The oscillations between different kinds of neutrino and
antineutrino flavor have been observed form solar, atmospheric
and LSND data. The three pieces of observation indicate three
values of mass-squared difference of three different orders. With
three families of neutrino, one can obtain only two independent
mass-squared differences. Therefore, observations require the
introduction of fourth neutrino which must be sterile in the
standard model. But many difficulties arise with the introduction
of fourth neutrino as discussed in literature (e.g. see
\cite{ggn}).

As an alternate proposal to accommodate the results, many authors
have proposed CPT violation in neutrino sector
\cite{bbls,bbl,my}. One can either introduce a new particle
(sterile neutrino) or allow CPT violation to take care of all
experimental results with present data. However, very recently,
MiniBooNE results have been declared which shows that LSND
experimental results can not be explained simply by neutrino
oscillation. Hence, it raises many other questions to be answered
\cite{mboo}.

Previously, many authors have explored the consequences of CPT
violation in neutrino sector (e.g. see \cite{bpww,bbbk}). The
nature of mass of neutrino has been studied \cite{bbbk} if CPT is
violated. It has been shown that if CPT is violated then
neutrinos no longer remain Majorana particles even if they have a
Majorana mass {\it i.e.} they violate lepton number. But the
actual physical condition which can lead to CPT violation was not
discussed satisfactorily. Proper situations in which CPT can be
violated were discussed later by many authors
\cite{mmp,sm,mpla,ahl1,ahl2,dmd,bmltst}. They have pointed out
that CPT violation in neutrino sector can occur due to the
spin-gravity coupling.

In the present paper, we plan to obtain the mass matrix for
neutrino sector with Majorana mass in the presence of
gravity. It has already been seen that spin gravity coupling has an extra
contribution to the effective mass of neutrino and antineutrino.
The Majorana type mass of neutrino alone can cause lepton number
violation {\it i.e.} the mixing of neutrino and antineutrino
states. Without an interaction with gravity, the neutrino-antineutrino
mixing angle is $\pi/4$. However, in presence of background gravitational field,
mixing angle changes and depends on the strength of CPT
violation. In a similar fashion, flavor oscillation is also altered by
gravitational effect. As a consequence of mixing, the neutrino-antineutrino
oscillation can take place only in presence of gravity. We also
show that gravity affects the neutrinoless double beta
decay rate even if two neutrino flavor mixing is taken.

We organize the paper in the following manner. In the next
section, we review the CPT violating nature of the spin-gravity
interaction for neutrino. In the section \ref{mmatrix}, we obtain
the mass matrix for neutrino traveling in a background
gravitational field. From that we find the neutrino-antineutrino
mixing and oscillation. Subsequently, we investigate effect of gravity on
flavor oscillation and its application to that in
early universe and around black holes in the section \ref{flmx}.
Applying this mixing, we recalculate the
neutrinoless double beta decay rate in the next section.
Finally, we summarize what we obtain and discuss their
implications in the section \ref{sum}.

\section{\label{gravity}Coupling to curvature}
The CPT violation mechanism due to the spin-gravity coupling of
fermions has been discussed earlier \cite{mmp,sm,mpla,dmd}.
How differently gravity interacts with neutrino than antineutrino
has been shown by a detailed calculation \cite{mpla,dmd,bmltst}.
For completeness, here we revisit the idea very briefly.

The general invariant coupling of spin-$1/2$ particles to
gravity is described by the Lagrangian \cite{par, hehl, fis,
chou, xia, lal, pal, waj, var} \begin{equation} {\cal
L}~=~\sqrt{-g}\left(\frac i2 \bar \Psi
\gamma^a\stackrel{\leftrightarrow}{D}_a\Psi- \bar \Psi m
\Psi\right), \label{lag} \end{equation} where the covariant
derivative is \begin{equation} D_a~=~\left(\partial_a-\frac i4
\omega_{bca}\sigma^{bc}\right) \end{equation} and the
spin-connections are
\begin{equation} \omega_{bca}~=~e_{b\lambda}\left(\partial_ae^\lambda_c+
\Gamma^\lambda_{\gamma\mu}e^\gamma_ce^\mu_a\right). \end{equation}
Here\begin{equation} \sigma^{bc}~=~\frac i2
\left[\gamma^b,\gamma^c\right]. \end{equation} All the above
equations have been written in a local inertial frame which is
flat along the entire geodesic. The Roman alphabets denote the
flat space indices, while the Greek alphabets denote the curved
space indices. Here $e$-s are the vierbeins connecting curved and
locally flat spaces and obey the relations $e^\mu_a e^{\nu
a}=g^{\mu \nu},~e^{a\mu}e^b_\mu=\eta^{ab}$, where $\eta^{ab}$
represents the inertial frame Minkowski metric and $g^{\mu \nu}$
the curved spacetime metric.

Thus the Lagrangian can be rewritten as

\begin{equation} {\cal L}~=~det(e)~\bar \Psi \left( \frac i2
\gamma^a\stackrel{\leftrightarrow}{\partial}_a-m+\gamma^a\gamma^5
B_a\right) \Psi, \label{eqn5} \end{equation} with \begin{equation}
B^d~=~\epsilon^{abcd}\omega_{bca}. \label{eqn6}
\end{equation}

In a local inertial frame, the effect of gravity on
fermion is only an axial vector interaction term involving the
gravitational field $B_a$ given by Eq. (\ref{eqn6}). The
Lagrangian given by Eq. (\ref{eqn5}) has two parts$-$ the free
part and the interaction part. The interaction part is a coupling
to the field $B_a$ which is constant in a local inertial frame
because that arises from the background gravitational field.

If $B_a$ is constant in a local inertial frame, then the
interaction term violates CPT as well as particle Lorentz
symmetry, although it is invariant under observer Lorentz
transformation (see e.g. \cite{coll}). It has been shown that
only a special form of the background gravitational field can
give rise to such CPT violating interaction term. For example, around
the rotating black holes \cite{sm,mpla}, or in anisotropic early
universe \cite{dmd}, a suitable background field exists which
causes CPT violation in interaction with spin-1/2 fermions.

The interaction term in the Lagrangian given by Eq. (\ref{eqn5})
involving $B_a$ contains $\gamma^5$. If we consider the standard model
neutrino with the Majorana mass term, then
right-handed neutrino and left handed antineutrino are absent. This
leads the gravitational interaction to $-\bar \Psi_L\gamma^a\Psi_L$ and $\bar
\Psi^c_L\gamma^a\Psi^c_L$ for neutrino and antineutrino
respectively. Here $c$ superscripted $\Psi$ represents the charge
conjugated spinor or the spinor for antiparticle and the subscript $L$
denotes the left handed component. Consequently, the
dispersion relations for neutrino and antineutrino become
different respectively given as
\begin{subequations}
\label{edis}
\begin{eqnarray}
%\nonumber
E_{\nu} =  \sqrt{({\vec p} - {\vec B})^2 + m^2} + B_0, \\
E_{\nu^c} = \sqrt{({\vec p} + {\vec B})^2 + m^2} - B_0
\end{eqnarray} with momentum $\vec{p}$.
\end{subequations}

\section{\label{mmatrix}Mass matrix}

If we consider Majorana neutrino, then in Weyl representation
neutrino spinor can be written as \begin{equation}
\Psi~=~\left(\begin{array}{c}\psi_L^c \\ \psi_L
\end{array}\right), \label{twocm}\end{equation} where $\psi_L^c$ and $\psi_L$ are two
component spinors for antineutrino and neutrino respectively
which are lepton number eigenstates with eigenvalues -1 and +1
respectively. Here we assume that neutrino is a left handed
particle only. Hence, hereafter we shall omit the subscript $L$.

In terms of two component spinors, the Majorana mass term can be
written as \begin{eqnarray} \bar \Psi^c M
\Psi~=~\left(\begin{array}{cc} \psi^{c\dag} & \psi^\dag
\end{array}\right)\left(\begin{array}{cc} 0 & -m
\\ -m & 0
\end{array}\right)\left(\begin{array}{c} \psi^c \\ \psi
\end{array}\right) \nonumber \\
=~-\psi^{c\dag} m \psi~-~\psi^\dag  m \psi^c. \label{mm}
\end{eqnarray}
Now in gravitational field the Lagrangian density can be written
as
\begin{widetext}
%\begin{eqnarray} (-g)^{-1/2} {\cal
%L}~=~\left(\begin{array}{cc} \psi^{c\dag} &
%\psi^\dag \end{array}\right)i \gamma^0 \gamma^\mu {\cal D}_\mu \left(\begin{array}{c}\psi^c \\
%\psi \end{array}\right)~+~\left(\begin{array}{cc} \psi^{c\dag} &
%\psi^\dag \end{array}\right)\gamma^5 B_0\left(\begin{array}{c}\psi^c \\
%\psi \end{array}\right)~-~\left(\begin{array}{cc} \psi^{c\dag} &
%\psi^\dag \end{array}\right)\gamma^0 m
%\left(\begin{array}{c}\psi^c
%\\ \psi \end{array}\right)\\~=~\left(\begin{array}{cc}
%\psi^{c\dag} &
%\psi^\dag \end{array}\right)i \gamma^0 \gamma^\mu {\cal D}_\mu \left(\begin{array}{c}\psi^c \\
%\psi \end{array}\right)~+~\left(\begin{array}{cc} \psi^{c\dag} &
%\psi^\dag \end{array}\right)\left(\begin{array}{cc}B_0 & 0 \\ 0
%& -B_0\end{array}\right)\left(\begin{array}{c}\psi^c \\
%\psi \end{array}\right)~-~\left(\begin{array}{cc}
%\psi^{c\dag} & \psi^\dag \end{array}\right)\left(\begin{array}{cc}
%0 & -m \\ -m & 0\end{array}\right)\left(\begin{array}{c}\psi^c
%\\ \psi \end{array}\right) \end{eqnarray}
\begin{eqnarray} (-g)^{-1/2} {\cal L}~=~\left(\begin{array}{cc}
\psi^{c\dag} & \psi^\dag \end{array}\right)\frac{i}{2} \gamma^0 \gamma^\mu
{\stackrel{\leftrightarrow}{\cal D}}_\mu \left(\begin{array}{c}\psi^c \\ \psi
\end{array}\right)~+~\left(\begin{array}{cc} \psi^{c\dag} &
\psi^\dag \end{array}\right)\gamma^5 B_0\left(\begin{array}{c}\psi^c \\
\psi \end{array}\right) \nonumber \\ ~-~\left(\begin{array}{cc}
\psi^{c\dag} & \psi^\dag \end{array}\right)\gamma^0 m
\left(\begin{array}{c}\psi^c
\\ \psi \end{array}\right) \nonumber \\  =~\left(\begin{array}{cc}
\psi^{c\dag} &
\psi^\dag \end{array}\right)\frac{i}{2} \gamma^0 \gamma^\mu \stackrel{\leftrightarrow}{\cal D}_\mu \left(\begin{array}{c}\psi^c \\
\psi \end{array}\right)~+~\left(\begin{array}{cc} \psi^{c\dag} &
\psi^\dag \end{array}\right)\left(\begin{array}{cc}B_0 & 0 \\ 0
& -B_0\end{array}\right)\left(\begin{array}{c}\psi^c \\
\psi \end{array}\right) \nonumber \\ ~-~\left(\begin{array}{cc}
\psi^{c\dag} & \psi^\dag \end{array}\right)\left(\begin{array}{cc}
0 & -m \\ -m & 0\end{array}\right)\left(\begin{array}{c}\psi^c
\\ \psi \end{array}\right) \end{eqnarray}
where \begin{equation} {\cal
D}_\mu~\equiv~(\partial_0,\partial_i+\gamma^5 B_i).
\end{equation}
Therefore, in Lagrangian density we obtain terms
containing $B_0 \psi^{c\dag}\psi^c$ and $B_0 \psi^\dag\psi$.
These two terms do not violate lepton number. Hence these terms
can be looked as lepton number non-violating {\it mass} of
Majorana neutrino. Therefore, writing the {\it mass} terms
together we obtain the Lagrangian density \begin{equation}
(-g)^{-1/2} {\cal L}~=~\left(\begin{array}{cc} \psi^{c\dag} &
\psi^\dag
\end{array}\right) \frac{i}{2} \gamma^0 \gamma^\mu \stackrel{\leftrightarrow}{\cal D}_\mu
\left(\begin{array}{c}\psi^c \\
\psi \end{array}\right)~-~\left(\begin{array}{cc}
\psi^{c\dag} & \psi^\dag \end{array}\right)\left(\begin{array}{cc}-B_0 & -m \\
-m & B_0\end{array}\right)\left(\begin{array}{c}\psi^c \\
\psi \end{array}\right) \end{equation} Hence, in background
gravitational field the Euler-Lagrange equation for neutrino and
antineutrino is
\begin{equation} i \gamma^0 \gamma^\mu {\cal D}_\mu \left(\begin{array}{c}\psi^c \\
\psi \end{array}\right)~-~\left(\begin{array}{cc}-B_0 & -m \\
-m & B_0\end{array}\right)\left(\begin{array}{c}\psi^c \\
\psi \end{array}\right)~=~0. \label{eleq}\end{equation}
\end{widetext}

Therefore, the mass matrix of the neutrino-antineutrino sector is
given by 
\begin{equation} {\cal M} =\left(\begin{array}{cc}-B_0 & -m \\
-m & B_0\end{array}\right). 
\label{fmm}
\end{equation} Hence we see that if
one assumes neutrino to have solely the Majorana type masses, then
it acquires lepton number non-violating
type masses, equal but opposite in sign, while propagating in a
gravitational field. Moreover, we see that in this case $\psi$ no longer
remains a mass eigenstate.

\subsection{Neutrino-antineutrino mixing and oscillation}

The mass matrix ${\cal M}$, given by Eq. (\ref{fmm}), is 
Hermitian and can be diagonalized by unitary transformation. Then the mass
eigenstates will be some admixture of $\psi$ and $\psi^c$.
Hence, neutrino and antineutrino states couple together.
We find these two mass eigenstates $\nu_1$
and $\nu_2$ as \begin{subequations}\begin{eqnarray} 
|\nu_1\rangle ~=~\frac1{N}\left\{\left(B_0+\sqrt{B_0^2+m^2}\right)
~|\psi^c\rangle~+~m~|\psi\rangle\right\} \\
|\nu_2\rangle~=~\frac1N\left\{-m~|\psi^c\rangle~+~\left(B_0+\sqrt{B_0^2+m^2}\right)
~|\psi\rangle\right\}.
\end{eqnarray} 
\label{st} 
\end{subequations} with eigenvalues \begin{equation} m_{1,2}~=
\mp\sqrt{B_0^2+m^2}.
\label{eqn16} \end{equation} Here \begin{equation}
N~=~\sqrt{2B_0^2+2m^2+2B_0\sqrt{B_0^2+m^2}} \end{equation} is the
normalization factor.  In a more convenient way, one can write
\begin{subequations}\label{mix}\begin{eqnarray}
|\nu_1\rangle~=~\cos \theta~|\psi^c\rangle~+~\sin
\theta~|\psi\rangle \\
|\nu_2\rangle~=~-\sin \theta~|\psi^c\rangle~+~\cos \theta~|\psi\rangle \end{eqnarray}
\end{subequations} with \begin{equation} \tan
\theta~=~\frac{m}{B_0+\sqrt{B_0^2+m^2}} \label{eqn20}.
\end{equation} Then one can write $|\psi^c\rangle$ and $|\psi\rangle$
as the superposition states of $|\nu_1\rangle$ and
$|\nu_2\rangle$ in the following manner
\begin{subequations}\label{eqn21}\begin{eqnarray}
|\psi^c\rangle~=~\cos \theta~|\nu_1\rangle~-~\sin \theta~|\nu_2\rangle \\
|\psi\rangle~=~\sin \theta~|\nu_1\rangle~+~\cos \theta~|\nu_2\rangle.
\end{eqnarray} \end{subequations}
Basically, $\nu_1=-i\,\sigma_2\,\nu_2^*$ as $\psi^c
=-i\,\sigma_2\,\psi$. Therefore, $\nu_1$ and $\nu_2$ together actually 
describe a single Majorana particle as $\psi$ and $\psi^c$ do.
Then, if we construct a new four component
neutrino spinor as
\begin{equation} \nu~\equiv~\left(\begin{array}{c}\nu_1 \\
                                                  \nu_2 \end{array}\right),
\label{eqn22}
\end{equation} 
it retains Majorana nature, {\it i.e.} $\nu^c~=~\nu$.
This is obvious as the new spinor $\nu$ is only a transformed
spinor from the previous one $\Psi$ and the old and new spinors
are related by a unitary transformation 
\begin{equation} 
\nu =\left(\begin{array}{cc} 
           \cos \theta & \sin \theta \\
           -\sin \theta & \cos \theta 
           \end{array}\right)~\Psi. 
\label{eqn26}
\end{equation}
Although, $\nu_1$ and $\nu_2$ are not lepton number
eigenstates, as evident from Eq. (\ref{st}), the mass terms in the
corresponding Lagrangian of $\nu_1$ and $\nu_2$ are lepton number
conserving. 
In absence of gravitational field, neutrino and antineutrino mix
in the same angle. 

Hence, if there is any lepton number violating interaction, then
we obtain two neutrino mass eigenstates $\nu_1$ and $\nu_2$ which
are superpositions of neutrino and antineutrino states given by
Eq. (\ref{mix}). As energies of neutrino and
antineutrino in gravitational field are different, there will be
an oscillation between $\nu_1$ and $\nu_2$.
At any time $t$, the oscillation probability is
given by \begin{equation} {\mathcal P}(t)~=~\sin^2 2\theta~\sin^2
\delta(t) \end{equation} where \begin{equation}
\delta(t)=\frac{|E_\nu- E_{\nu^c}|t}2, \end{equation} when $E_\nu$
and $E_{\nu^c}$ are given by Eq. (\ref{edis}). Hence, for ultra
relativistic neutrinos by putting the value of $\theta$, we obtain
\begin{equation} {\mathcal P}(t)~=~\frac{m^2}{B_0^2+m^2}~\sin^2 \{(B_0-|\vec
B|)t\} \label {oscprob} \end{equation} for neutrino and
antineutrino of same rest mass $m$. Thus the oscillation length is
\begin{equation} \lambda=\frac{\pi}{B_0-|\vec B|}. \end{equation} This depends only
on the strength of the gravitational field. If we consider
neutrinos to be coming out off the inner accretion disk around a
spinning black hole of mass $M=10M_\odot$, then
$B_0-|\vec{B}|=\tilde{B}=10^{-19}$ GeV \cite{bmltst}, which leads
to $\lambda\sim 10$km. If the disk is around a supermassive black
hole of $M=10^8M_\odot$ in an AGN, then $\lambda$ may increase
to $10^{10}$km, depending upon the size of inner edge where from
neutrinos come out and angular momentum of the black hole.
Therefore, an oscillation may complete from a few factor to
hundred Schwarzschild radii in the disk producing copious
antineutrino over neutrino and may cause overabundance of neutron
and positron. However, neutrinos around a primordial black hole
of mass $M_p$ \footnote{Note that the corresponding temperature
$T_p\sim 10^{-20}\,M_\odot/M_p$ GeV \cite{st}.} could lead to an
oscillation length as small as $\lambda\sim 10^{-16} {\rm km} \le 100M_p$ for
$r\le 100M_p$.

Therefore, from Eqs. (\ref{mix}) and (\ref{eqn20}) we see that for
nonzero value of Majorana mass ($m\neq0$), neutrino and
antineutrino combine to give two new states $\nu_1$ and $\nu_2$,
mass eigenstates with different mass eigenvalues. Therefore, in
these two states neutrino and antineutrino are mixed. In the
absence of gravity, the mixing angle is $\pi/4$ which is evident
from the Eq. (\ref{eqn20}). On the other hand, if
$m=0$, {\it i.e.} there is no lepton number violating interaction, then
neutrino and antineutrino do not couple at all [see Eq.
(\ref{eleq})]. This means two component neutrino and antineutrino described
in Eq. (\ref{twocm}) themselves
are mass eigenstates. In this case, presence of gravity which is
CPT violating, splits these two eigenstates with two different
mass eigenvalues.

We see that although initially gravitational interaction and the
Majorana mass term explicitly have different effects, one to
violate CPT and another to violate lepton number, both of them
contribute in the same manner to split the mass eigenstates (see
Eq. (\ref{eqn16})). Moreover, in gravitational field, since
neutrino and antineutrino acquire different effective masses,
gravitational field coupled to neutrino spin may have some lepton
number violating nature implicitly. This has been illustrated in
literature \cite{mpla,dmd}. We also see from Eq. (\ref{oscprob})
that presence of gravity leads to oscillation. Without
gravitational field, the lepton number violating interaction
alone can not cause this oscillation.

\subsection{Oscillation with lepton number conserving mass}

For Majorana neutrino when the sterile components are neglected,
the Dirac mass is of no relevance. This is exactly the case we
are considering in our present paper. We have also seen that the
gravity can induce an effective mass for Majorana neutrino which
does not violate lepton number. If we consider the diagonal term
of the mass matrix of Eq. (\ref{mm}) to be non-zero, then we
obtain an extra mass term which conserves the lepton number. With
this mass term included, the effective mass matrix will
take the form \begin{equation} {\cal M}_n =\left(\begin{array}{cc}m_n-B_0 & -m \\
-m & m_n+B_0\end{array}\right), \end{equation} where $m_n$ is the
so called lepton number non-violating mass.

With this mass matrix, the mixing angle of neutrino and
antineutrino does not alter. However, the mass eigenvalues of the
mass eigenstates become \begin{equation} m_{n(1,2)}~=
m_n\mp\sqrt{B_0^2+m^2} \label{eqn31}. \end{equation}

In this case, we can consider oscillation between neutrino and
antineutrino as they are combination of $\nu_1$ and $\nu_2$ which
evolve differently with time. For this oscillation, the
oscillation probability at any time $t$ is given by
\begin{equation} {\mathcal P}_n(t)~=~\sin^2 2\theta~\sin^2 \epsilon(t)
\end{equation} where \begin{equation} \epsilon(t)=\frac{|E_1-
E_2|t}2, \end{equation} when $E_1$ and $E_2$ are energies of two
neutrino mass eigenstates with momentum $\vec{p}$. In the
ultra-relativistic limit ($|\vec{p}|>>m$) \begin{equation}
E_{(1,2)}~=~|\vec{p}|+\frac{m_{n(1,2)}^2}{2|\vec{p}|}.
\end{equation} Again assuming $|\vec{p}|\sim E$ \begin{equation}
E_2-E_1=\frac{m_{n2}^2-m_{n1}^2}{2E} =\frac{2
m_n\sqrt{B_0^2+m^2}}{E}. \end{equation} Hence the the oscillation
probability at any time $t$ is given by \begin{equation} {\mathcal
P}_n(t)~=~\frac{m^2}{B_0^2+m^2}~\sin^2
\left(\frac{m_n\sqrt{B_0^2+m^2}}{E}t\right) \label {oscprobn}
\end{equation} and the oscillation length is given by \begin{equation} \lambda_n=\frac{ \pi
E}{m_n\sqrt{B_0^2+m^2}}. \end{equation}

Hence we see that this neutrino-antineutrino oscillation
probability does not depend on the spatial part of the
gravitational field (gravitational vector potential) $\vec{B}$
but depends on the temporal part (gravitational scalar potential)
$B_0$ only. However, there is nothing new in it. It was already
shown that neutrino asymmetry and then leptogenesis in early
universe arises due to non-zero $B_0$ \cite{dmd} independent of
$\vec{B}$. The gravitational scalar potential $B_0$ is non-zero
if the background spacetime has at least one nonzero off-diagonal
spatial component when the set of coordinate variables is
$\{t,x,y,z\}$. Existence of such a component in a spacetime may
be due to anisotropy which is the case for the Bianchi model.
Moreover, off-diagonal spatial components also govern in presence
of primordial quantum fluctuations in the  Robertson-Walker
spacetime. This is basically the tensor perturbation to early
universe. Therefore, it is confirmed that the gravity induced
leptogenesis is possible
only if the gravitational scalar potential is non-zero.

\section{\label{flmx}Effect of gravity on flavor mixing and oscillation}

\subsection{Flavor mixing}

In previous section we have seen that interaction of neutrino
and antineutrino with gravity gives rise to neutrino mass
eigenstates which are superposition of neutrino and antineutrino
states given by
\begin{equation} 
\nu_{e,\mu}~=~{\cal U}_{e,\mu}^\dag \Psi_{e,\mu}
\label{matqg2}
\end{equation}
where
%in the Appendix \ref{App:anti}.
\begin{equation} {\cal U}_{e,\mu}~=~\left(
   \begin{array}{cc}
    \cos \theta_{e,\mu} & -\sin \theta_{e,\mu} \\

    \sin \theta_{e,\mu} & \cos \theta_{e,\mu} \\ \end{array}
   \right).
    \end{equation}

It is clear from our above discussions that $\nu_{e,\mu}$ here are four
component spinors. In other way, it can be stated that in gravitational field
neutrino state $\Psi$ of mass $m$ is modified to the state $\nu$ of 
mass $\sqrt{B_0^2+m^2}$ as described in section \ref{mmatrix}.
Therefore, recalling Eq. (\ref{eqn22}) we construct flavor eigenstates
$\nu_e$ and $\nu_\mu$ under gravity for electron and muon neutrino respectively
whose components are $\nu_{e1}$, $\nu_{e2}$, $\nu_{\mu1}$ and $\nu_{\mu2}$
with masses $m_{e1}$, $m_{e2}$, $m_{\mu1}$ and $m_{\mu2}$ respectively. In
terms of gravitational coupling these mass eigenvalues are expressed as
\begin{eqnarray}
m_{(e,\mu)1}~=~-\sqrt{B_0^2+m_{e, \mu}^2}, \nonumber \\
m_{(e,\mu)2}~=~\sqrt{B_0^2+m_{e, \mu}^2},
\label{flvmas}
\end{eqnarray}
where $m_e$, $m_\mu$ are the Majorana masses for electron and muon neutrino
respectively, analogous to $m$ of section III.
The corresponding mixing parameters are given by
%in Appendix \ref{App:anti}.
\begin{equation}
\tan \theta_{e,\mu}~=~\frac{m_{e, \mu}}{B_0+\sqrt{B_0^2+m_{e,
\mu}^2}}. \label{tant}
\end{equation}

Now we consider the corresponding two flavor mixing.
The states $\nu_e$ and $\nu_\mu$ are coupled by a
Majorana mass term $m_{e\mu}$. Then the mass term in the
Lagrangian density is considered as
\begin{widetext}
\begin{eqnarray} \nonumber (-g)^{-1/2} {\cal
L}_m~&=&~-\frac12\left(\nu_{e1}^\dag
m_{e1}\nu_{e1}~+~\nu_{e2}^\dag m_{e2}\nu_{e2}~+~\nu_{\mu1}^\dag
m_{\mu1}\nu_{\mu1}~+~\nu_{\mu2}^\dag
m_{\mu2}\nu_{\mu2}\right.\\&-& \left.~\nu_{\mu1}^\dag m_{e\mu}
\nu_{e1}~-~\nu_{e1}^\dag m_{e\mu}\nu_{\mu 1}~+~\nu_{\mu2}^\dag
m_{e\mu} \nu_{e2}~+~\nu_{e2}^\dag m_{e\mu}
\nu_{\mu2}\right).
\label{lagmass}
\end{eqnarray}
\end{widetext}
Here we assume that the Majorana mass matrix, coupling $\nu_e$ and
$\nu_\mu$, is Hermitian and diagonal. We also assume that $m_{e\mu}$
in $\nu^\dag_{e1}\nu_{\mu1}$ and $\nu^\dag_{e2}\nu_{\mu2}$ are
same for computational simplicity, while our main goal is to
investigate any gravity effect. As our main aim is to study the effect of
curvature to the oscillation phase, even for the
convenience of transparent understanding of sole effect of gravity we
prefer to consider the nongravitating part as simple as possible. 
This mass term gives rise
to two mass matrices ${\cal M}_{f(1,2)}$ given by

\begin{equation}
{\cal M}_{f1}~=~\left(\begin{array}{cc}
                      m_{e1} & -m_{e\mu} \\
                      -m_{e\mu} & m_{\mu1}
                      \end{array}\right),~~~~~~~
{\cal M}_{f2}~=~\left(\begin{array}{cc}
                      m_{e2} & m_{e\mu} \\
                      m_{e\mu} & m_{\mu2}

                      \end{array}\right)
\end{equation}
which mix $\nu_{e1}$ ($\nu_{e2}$) and $\nu_{\mu 1}$ ($\nu_{\mu 2}$). 
The mixing, leads to two mass eigenstates $f_{11}$ ($f_{12}$) and $f_{21}$
($f_{22}$), is expressed as
\begin{equation}
f~=~ {\cal F}^\dag \nu_f
\label{matqf}
\end{equation}
where
\begin{equation}
\nu_f~=~\left(
\begin{array}{c}
\nu_e\\
\nu_\mu \\
\end{array}
\right),~~~ f~=~\left(
\begin{array}{c}
f_1 \\
f_2 \\
\end{array}
\right) \end{equation}
and ${\cal F}$ is given by
\begin{equation} {\cal F}_{1,2}~=~\left(
   \begin{array}{cc}
    \cos \phi_{1,2} & -\sin \phi_{1,2} \\

    \sin \phi_{1,2} & \cos \phi_{1,2} \\ \end{array}
   \right).
    \end{equation}
Here the subscript $1$ and $2$ refer to the flavor mixing between mass eigenstates subscripted by
$1$ and $2$ respectively.
%Here the flavor mixing angle $\phi$ is
%different for states with subscript $1$ and $2$.
Hence
\begin{equation} \tan \phi_{1,2}~=~\frac{\mp 2m_{e\mu}}{(m_{e(1,2)}-m_{\mu(1,2)})+\sqrt{(m_{e(1,2)}-m_{\mu(1,2)})^2+
4m_{e\mu}^2}}.
\end{equation}
Thus we obtain all together four mass eigenstates $\chi_1$, $\chi_2$, $\chi_3$ and
$\chi_4$ described as
\begin{equation} \left(\begin{array}{c} \chi_1 \\ \chi_2 \\
\end{array} \right)~\equiv~\left(\begin{array}{c} f_{11} \\ f_{21} \\
\end{array} \right)~=~{\cal F}^\dag_1 \left(\begin{array}{c} \nu_{e1} \\
\nu_{\mu1}
\end{array} \right)~~~{\rm and}~~~\left(\begin{array}{c} \chi_3 \\ \chi_4 \\
\end{array} \right)~\equiv~\left(\begin{array}{c} f_{12} \\ f_{22} \\
\end{array} \right)~=~{\cal F}^\dag_2 \left(\begin{array}{c} \nu_{e2} \\
\nu_{\mu2}
\end{array} \right) \end{equation}
with mass eigenvalues $M_1$, $M_2$, $M_3$ and $M_4$ given by

\begin{eqnarray}
M_{1,2}~\equiv~m_{(1,2)1}~=~\frac 12 \left\{(m_{e1}+m_{\mu 1})\pm
\sqrt{(m_{e1}-m_{\mu 1 })^2+4m_{e \mu}^2} \right\} \nonumber \\
M_{3,4}~\equiv~m_{(1,2)2}~=~\frac 12 \left\{(m_{e2}+m_{\mu 2})\pm
\sqrt{(m_{e2}-m_{\mu 2 })^2+4m_{e \mu}^2} \right\}.
\end{eqnarray}
Now from Eq. (\ref{flvmas})

\begin{eqnarray}
m_{e1}+m_{\mu
1}~=~-\left(\sqrt{B_0^2+m_e^2}+\sqrt{B_0^2+m_\mu^2}\right)~=~-(Y_e+Y_\mu)
\nonumber \\
m_{e2}+m_{\mu
2}~=~\left(\sqrt{B_0^2+m_e^2}+\sqrt{B_0^2+m_\mu^2}\right)~=~Y_e+Y_\mu
\nonumber \\
m_{e1}-m_{\mu
1}~=~-\sqrt{B_0^2+m_e^2}+\sqrt{B_0^2+m_\mu^2}~=~-Y_e+Y_\mu
\nonumber \\
m_{e2}-m_{\mu
2}~=~\sqrt{B_0^2+m_e^2}-\sqrt{B_0^2+m_\mu^2}~=~Y_e-Y_\mu
\label{ydef}
\end{eqnarray}
where
\begin{equation}
Y=\sqrt{B_0^2+m^2}.
\label{yy}
\end{equation}
If we define
\begin{equation}
2m_a~=~Y_\mu+Y_e,~~~~~2m_i~=~Y_\mu-Y_e \label{mdef}
\end{equation}
then
\begin{eqnarray}
m_{e1}+m_{\mu1}=-2m_a,~~~~~m_{e1}-m_{\mu1}=2m_i, \nonumber \\
m_{e2}+m_{\mu2}=2m_a,~~~~~m_{e2}-m_{\mu2}=-2m_i.
\end{eqnarray}
Then we have
\begin{eqnarray}
M_1~=~-m_a~+~\sqrt{m_i^2+m_{e\mu}^2},~~~~~~~~
M_2~=~-m_a~-~\sqrt{m_i^2+m_{e\mu}^2}, \nonumber \\
M_3~=~m_a~+~\sqrt{m_i^2+m_{e\mu}^2},~~~~~~~~~
M_4~=~m_a~-~\sqrt{m_i^2+m_{e\mu}^2}, \label{mas4}
\end{eqnarray}
and
\begin{equation}
\tan \phi_{1,2}~=~\frac{\mp m_{e\mu}}{\pm m_i~+~\sqrt{m_i^2+m_{e\mu}^2}}. \label{tana}
\end{equation}
Note that at $B_0=0$ all the above results reduce to that of conventional two flavor
mixing without gravity effect. For example, at $B_0=0$ 
\begin{eqnarray}
\nonumber
\psi_{e,\mu}=\frac{1}{\sqrt{2}}(\nu_{(e,\mu)1}+\nu_{(e,\mu)2}),\\
\psi_{e,\mu}^c=\frac{1}{\sqrt{2}}(\nu_{(e,\mu)1}-\nu_{(e,\mu)2}).
\end{eqnarray}
Then substituting $\nu$-s by $\psi$-s in Eq. (\ref{lagmass}) with
$m_{(e,\mu)1}=-m_{e,\mu}$ and $m_{(e,\mu)2}=m_{e,\mu}$ one obtains
the standard mass Lagrangian density with spinors $\psi$ and $\psi^c$ 
\cite{bilpon}
\begin{eqnarray}
%\nonumber
{\cal L}_m=-\frac{1}{2}({\psi_e^c}^\dag m_e \psi_e + {\psi_e^c}^\dag m_{e\mu}
\psi_\mu + {\psi_\mu^c}^\dag m_{e\mu} \psi_e +{\psi_\mu^c}^\dag m_\mu 
\psi_\mu)+h.c.
\label{lagmass0}
\end{eqnarray}
The readers comparing this result with that in \cite{bilpon} should
not confuse with difference in notation in the present paper from 
that in \cite{bilpon}. Here $\nu_{e,\mu}$ are the 
Majorana neutrino fields of electron and muon typed under gravity
whose oscillation would be interesting to study.

%\subsection{\label{just}Justification of assumption }

The most general flavor mixing mass matrix in presence of gravity is
given by 
%\begin{displaymath}
\begin{eqnarray}
{\cal M}_4=\left(
\begin{array}{cc}
-B_0 {\mathbf I} & -{\mathbf M}\\
-{\mathbf M} & B_0 {\mathbf I} \\
\end{array}
\right),
\end{eqnarray}
%\end{displaymath}
when ${\mathbf I}$ is the $2\times 2$ unit matrix and
\begin{eqnarray}
%\begin{displaymath}
{\mathbf M}=
\left(
\begin{array}{cc}
m_e & m_{e\mu}\\
m_{e\mu} & m_\mu
\end{array}\right)\equiv {\mathbf U}_\theta\cdot {\rm diag}(m_1, m_2)
\cdot{\mathbf U}_\theta^\dag,
%\end{displaymath}
\end{eqnarray}
where ${\mathbf U}_\theta$ and $m_{1,2}$ are the mixing matrix in vacuum
and the neutrino masses respectively in absence of curvature.
The the corresponding flavor state is given by
\begin{eqnarray}
%\begin{displaymath}
\Psi=\left(
\begin{array}{c}
\psi^c_e\\
\psi^c_\mu\\
\psi_e\\
\psi_\mu\\
\end{array}
\right).
%\end{displaymath}
\end{eqnarray}
However, for the sake of simplicity we
take an assumption such that the $4\times4$ mass matrix is
resolved into $2\times2$ block form. The motivation is to study the
gravity effect in a transparent manner. In order to do that we consider
the mixing part by part as described above: 
first we take the gravity effect on the flavor eigenstates and
then we mix them up. The argument for this assumption is as follows. We
detect neutrinos by weak interactions.
In standard experiments, we detect and produce them via
charge-current interaction. As the flavor eigenstates
take part in charge-current interaction, we detect and produce
them only. In our calculation, the original flavor eigenstates are $\psi$
and $\psi^c$ or as a whole $\Psi$. Therefore, we plan to discuss
the oscillation probability and corresponding length among states $\Psi$.
However, when we consider them
under strong gravity, they couple and no longer have been a definite
mass state, rather modify to a state $\nu$. Hence it is worthwhile to
study the oscillation properties for states $\nu$ as well. Indeed the solution
of the Dirac equation in presence of gravitational
interaction is different than that in absence of gravity. 
Hence while performing the oscillation experiment
under gravity, it can be assumed that
gravity will affect the initially produced flavor states and in between the
detection and production the flavor mixing Hamiltonian will act upon these
gravitationally modified flavor states. Therefore, even though they are produced
as flavor eigenstates $\Psi$ via weak interaction, if system
is under strong gravity, then they are modified to $\nu$, as described
in \S III.A,B, which 
may be observable in presence of strong curvature.
Indeed $\nu\equiv\Psi$ when $B_0=0$. 
%\underline {We argue that in presence
%of gravity particle and antiparticle couple each other and thus
%the 4 component spinor, what the referee mentions, does not exist.}
If we put $B_0= 0$ in our oscillation probability and oscillation length, 
described in \S IV.B below,
then we obtain the standard expressions of 
those in flat space. This validates our assumption of mixing
scheme and verifies that our results
for $\nu$ are physically same as that for $\Psi$ under gravity, 
as will be discussed in \S IV.C.

\subsection{\label{flos}Flavor oscillation}

The sets of two Majorana neutrino flavor eigenstates are described as
\begin{subequations}\label{fl1}\begin{eqnarray} |\nu_{e1}
\rangle~=~\cos{\phi_1}|\chi_1\rangle~-~\sin{\phi_1}|\chi_2\rangle \\
|\nu_{\mu1}\rangle~=~\sin{\phi_1}|\chi_1\rangle~+~\cos{\phi_1}|\chi_2\rangle
\end{eqnarray}\end{subequations} and \begin{subequations}\label{fl2}\begin{eqnarray} |\nu_{e2}
\rangle~=~\cos{\phi_2}|\chi_3\rangle~-~\sin{\phi_2}|\chi_4\rangle \\
|\nu_{\mu2}\rangle~=~\sin{\phi_2}|\chi_3\rangle~+~\cos{\phi_2}|\chi_4\rangle.
\end{eqnarray}\end{subequations}

Let us now consider the states given by Eqs. (\ref{fl1}). The
oscillation probability of those states is \begin{equation} {\cal
P}_{fg1}~=~\sin^22\phi_1\sin^2\delta_{fg1}(t), \end{equation} where
\begin{equation} \delta_{fg1}(t)~=~\frac{|M_1^2-M_2^2|}{4E}~t \end{equation}
and $E$ is the energy of the mass eigenstates. Hence the
oscillation length is
\begin{equation} \lambda_{fg1}~=~ \frac{4\pi E}{|M_1^2-M_2^2|}.
\end{equation}
Similarly, for the states of Eqs. (\ref{fl2}) the
oscillation probability is \begin{equation} {\cal
P}_{fg2}~=~\sin^22\phi_2\sin^2\delta_{fg2}(t), \end{equation}  where
\begin{equation} \delta_{fg2}(t)~=~\frac{|M_3^2-M_4^2|}{4E}~t \end{equation}
and the oscillation length is \begin{equation} \lambda_{fg2}~=~ \frac{4\pi E}{|M_3^2-M_4^2|}.
\end{equation}
Now from Eqs. (\ref{ydef}), (\ref{mdef}) and (\ref{mas4}) we
obtain
\begin{eqnarray}
\Delta M^2=|M_1^2-M_2^2|~=~|M_3^2-M_4^2| \nonumber \\
=~\left(\sqrt{B_0^2+m_\mu^2}+\sqrt{B_0^2+m_e^2}\right) \nonumber
\\
\times\sqrt{\left\{\left(\sqrt{B_0^2+m_\mu^2}-\sqrt{B_0^2+m_e^2}\right)^2+4m_{e\mu}^2\right\}}.
\end{eqnarray}
From Eq. (\ref{tana}), we obtain
\begin{equation} \sin^2 2\phi_1~=~\sin^2 2\phi_2~=~\sin^2 2\phi=\frac{m_{e\mu}^2}
{m_i^2+m_{e\mu}^2}\label{sina}
\end{equation}
so that the oscillation probability and the oscillation length
are same for the two cases respectively given by
\begin{equation} {\cal P}_{fg}~=~\sin^2 2\phi~\sin^2 \delta_{fg}(t),
\label{ospgen}
\end{equation}
\begin{equation} \lambda_{fg}~=~\frac{4\pi E}{\Delta M^2}.
\label{oslgen}
\end{equation}

However, $\Psi$-s are related to $\nu$-s. Hence once we obtain
the oscillation probability and length between $\nu_{e1}$ (or $\nu_{e2}$) 
and $\nu_{\mu1}$ (or $\nu_{\mu2}$), the
results can be converted to that for $\psi_e$ and $\psi_\mu$ given in the
next subsection.

\subsection{\label{oscpsi}Oscillation probability between initial flavors
produced via weak interaction}

It is evident from Eqs. (\ref{matqg2}) and (\ref{matqf}) that we can express the initial
flavor states $\Psi_e$ and $\Psi_\mu$ in terms of mass eigenstates $\chi$-s
given by
\begin{eqnarray}
\psi_e^c~=~\cos \theta_e~\cos \phi_1~ \chi_1~-~\cos \theta_e~\sin \phi_1 ~\chi_2
~-~\sin \theta_e~\cos \phi_2 ~\chi_3~+~\sin \theta_e~\sin \phi_2 ~\chi_4,
\nonumber \\
\psi_\mu^c~=~\cos \theta_\mu~\sin \phi_1~ \chi_1~+~\cos \theta_\mu~\cos \phi_1
~\chi_2 ~-~\sin \theta_\mu~\sin \phi_2 ~\chi_3~-~\sin \theta_\mu~\cos \phi_2
~\chi_4, \nonumber \\
\psi_e~=~\sin \theta_e~\cos \phi_1 ~\chi_1~-~\sin \theta_e~\sin \phi_1 ~\chi_2
~+~\cos \theta_e~\cos \phi_2 ~\chi_3~-~\cos \theta_e~\sin \phi_2 ~\chi_4,
\nonumber \\
\psi_\mu~=~\sin \theta_\mu~\sin \phi_1 ~\chi_1~+~\sin \theta_\mu~\cos \phi_1
~\chi_2 ~+~\cos \theta_\mu~\sin \phi_2 ~\chi_3~+~\cos \theta_\mu~\cos \phi_2
~\chi_4.
\label {Imixstates}
\end{eqnarray}
In short, we can write this as
\begin{equation} 
\left( \begin{array}{c}
        \psi^c_e \\
        \psi^c_\mu \\
        \psi_e \\
        \psi_\mu 
\end{array} \right)~=~T\left( \begin{array}{c}
                             \chi_1 \\
                             \chi_2 \\
                             \chi_3 \\
                             \chi_4 
\end{array} \right),  
\end{equation}
where

\begin{widetext}

\begin{equation} 
T~=~\left( \begin{array}{cccc}
            \cos \theta_e \cos \phi_1 & -\cos \theta_e \sin \phi_1 & 
            -\sin \theta_e \cos \phi_2 & \sin \theta_e \sin \phi_2 \\

            \cos \theta_\mu \sin \phi_1 & \cos \theta_\mu \cos \phi_1 & 
            -\sin \theta_\mu \sin \phi_2 & -\sin \theta_\mu \cos \phi_2  \\

            \sin \theta_e \cos \phi_1 & -\sin \theta_e \sin \phi_1 & 
            \cos \theta_e \cos \phi_2 & -\cos \theta_e \sin \phi_2 \\

            \sin \theta_\mu \sin \phi_1 & \sin \theta_\mu \cos \phi_1 & 
            \cos \theta_\mu \sin \phi_2 & \cos \theta_\mu \cos \phi_2 \\ 
            \end{array} \right).
\end{equation} 

\end{widetext}
From this the particle part $\psi$ can be written as 
\begin{eqnarray}
\psi_e~=~T_{e1} ~\chi_1~+~T_{e2} ~\chi_2
~+~T_{e3}~\chi_3~+~ T_{e4}~\chi_4,
\nonumber \\
\psi_\mu~=~T_{\mu 1} ~\chi_1~+~ T_{\mu 2}
~\chi_2 ~+~T_{\mu 3} ~\chi_3~+~T_{\mu 4}
~\chi_4.
\label {symbpart}
\end{eqnarray}
Hence we compute the oscillation probability between $\psi_e$ and $\psi_\mu$  
\begin{equation}
{\cal P}_{ig}=-4\sum_{i<j=1}^4T_{e i }T_{e j}T_{\mu i}T_{\mu j}
\sin^2 \delta_{ij}(t),
\label{oscprobpsi}
\end{equation}
where 
\begin{equation}
\delta_{ij}(t)=\frac{|\Delta M^2_{ij}|}{4E}t,~~~~\Delta M^2_{ij}=M_i^2-M_j^2.
\label{deltai}
\end{equation}
This resembles the oscillation probability one obtains 
in the case of 3 flavors mixing or in general the case of $N$ flavors mixing.
With actual expressions of $T_{(e,\mu)i}$-s from Eq. (\ref{Imixstates})
we obtain 
\begin{eqnarray}
{\cal P}_{ig}=~\sin^2 \theta_e
~\sin^2 \theta_{\mu}~\sin^2 2 \phi_1~\sin^2 \delta_{12}(t)
~+~\cos^2 \theta_e~\cos^2 \theta_{\mu}~\sin^2 2 \phi_2~\sin^2 \delta_{34}(t)
\nonumber \\
-~\frac 14 \sin 2 \theta_e~\sin 2 \theta_\mu ~\sin 2 \phi_1
~\sin 2 \phi_2 
\left\{\sin^2\delta_{13}(t)-\sin^2\delta_{14}(t)
-\sin^2 \delta_{23}{t}+\sin^2 \delta_{24}(t)
\right\}.  
\label{oscprobgi}
\end{eqnarray}
From Eq. (\ref{mas4})
\begin{equation}
|M_1^2-M_3^2|~=~|M_2^2-M_4^2|~=~\Delta M^2~=~4 m_a \sqrt{m_i^2+m_{e\mu}^2}
\label{delmsq1}
\end{equation}
and
\begin{equation}
|M_1^2-M_4^2|~=~|M_2^2-M_3^2|~=~0.
\label{delmsq0}
\end{equation}
Moreover, from Eq. (\ref{tana}) 
\begin{equation}
\sin 2\phi_1~=~-\sin 2\phi_2~=~\sin 2\phi~ ({\rm say}).
\end{equation}
Hence the oscillation probability is given by 
\begin{equation}
{\cal P}_{ig}~=~ \sin^2 2 \phi \left\{(\sin \theta_e \sin \theta_\mu 
+ \cos \theta_e \cos \theta_\mu)^2 \sin^2 \left(\frac{\Delta M^2}{4E}t\right) 
\right\}.
\end{equation}
Therefore, we find that the oscillation length between $\psi$ states is same 
as that in the case of oscillation between $\nu_1$ (and $\nu_2$) states:
equivalent to each other.

Now the oscillation length would change for a distant observer
due to gravitational redshift. Therefore, the redshifted oscillation length
\begin{equation}
\lambda_{fr}~=~\frac{\lambda_f}{\sqrt{g_{tt}}}. \end{equation} If
we consider the neutrino flavor oscillation in a
rotating black hole spacetime of mass $M$ and specific angular
momentum $a$, then at a point $(r,\theta)$
\begin{equation}
g_{tt}~=~1-\frac{2Mr}{\rho^2} \end{equation} with
\begin{equation} \rho^2~=~r^2+a^2\cos \theta. \end{equation}
Therefore, it is evident from the above discussions that under gravity the probability of
flavor oscillation changes significantly depending on the gravitational strength.

We now consider Majorana mass of electron neutrino $m_e\sim5\times 10^{-3}$ eV,
muon neutrino $m_\mu\sim6.5\times 10^{-3}$ eV and mixing $m_{e\mu}\sim 3.5\times
10^{-3}$ eV. These values are consistent with the solar neutrino oscillation
data.  Thus, without gravity the
oscillation probability between electron and muon neutrino with
$\Delta M^2(B_0=0)=8.2\times 10^{-5}\,{\rm eV^2}$,
%and mixing angle $30^o <\phi(B_0=0) <38^o$,
as from solar neutrino data \cite{alex}, is given by
\begin{equation}
{\cal P}_f={\cal P}_{fg}(B^0=0)~\simeq~0.956~\sin^2\left(\frac{8.2\times10^{-5}~{\rm eV^2}}{4E}~t\right).
\label{ospf}
\end{equation}
The corresponding oscillation length
\begin{equation}
\lambda_f=\lambda_{fg}(B_0=0)\simeq \frac{4\pi E}{8.2\times10^{-5}{\rm eV^2}}.
\label{oslf}
\end{equation}

\subsection{Oscillation around black holes}

We first recall the gravitational scalar potential computed earlier in the Kerr geometry \cite{bmltst}
\begin{equation}
B^0=-\frac{4 a\sqrt{M} z}{{\bar{\rho}}^2\sqrt{2 r^3}}
\label{b0ker}
\end{equation}
for a black hole of mass $M$ and specific angular momentum $a$, where
${\bar{\rho}}^2 = 2r^2+a^2-x^2-y^2-z^2$.
If we consider neutrinos at around $20$ Schwarzschild radius in the spacetime of a
primordial black hole of mass $10^{22}$gm, then from Eq. (\ref{b0ker})
the gravitational field is computed to be $\sim10^{-2}~{\rm eV}$
which is comparable to neutrino masses.
If we specify the gravitational field $B_0\sim 5\times 10^{-2}$ eV, then
from Eqs. (\ref{ospgen}) the oscillation probability becomes
\begin{equation} {\cal P}_{fBH}~\simeq~0.999~\sin^2\left(\frac{7\times10^{-4}\,{\rm eV^2}}{4E}~t\right).
\label{ospbh}
\end{equation}
The oscillation length, from Eq. (\ref{oslgen}), is determined as
\begin{equation}
\lambda_{fBH}\simeq \frac{4\pi E}{7\times10^{-4}~{\rm eV^2}}.
\label{oslbh}
\end{equation}
Hence, the oscillation length decreases by almost an order of magnitude.

If $B_0$ is much higher than neutrino Majorana masses, then from the
Eq. (\ref{sina}) we have $\sin^22\phi\sim 1$ and $\Delta M^2\sim 4m_{e\mu}B_0$.
Therefore, the oscillation probability
\begin {equation}
{\cal P}_{fBH}=\sin^2 \left(\frac{1.4\times10^{-2}\,{\rm eV} B_0}{4E}\, t\right)
\end{equation}
and the oscillation length
\begin{equation}
\lambda_{fBH}
%\simeq\frac{2\pi E}{4m_{e\mu}B_0}
%\simeq\frac{5\times 10^2 \pi E}{B_0}{\rm eV^{-1}}
\simeq\frac{4 \pi E}{1.4\times 10^{-2}\,{\rm eV} B_0}.
\label{oslbhh}
\end{equation}

\subsection{Oscillation in early universe}
\subsubsection{Anisotropic universe}
We recall the gravitational scalar potential in the anisotropic phase of
early universe \cite{dmd,bmltst}
\begin{eqnarray}
B^0=\frac{4R^3S+3y^2R\,S^3-2y\,S^4}{8R^4+2y^2R^2S^2}.
\label{bobian}
\end{eqnarray}
If we consider radiation dominated era with $R(t)=(t/t_0)^{1/2}$ and $S(t)$ as an
arbitrary constant $\ge 1$, then the above potential reduces to
\begin{eqnarray}
B_0\sim\frac{S^2}{y}\left(\frac{t_0}{t}\right),
\label{bobiansp}
\end{eqnarray}
when $y$ is the position coordinate, can not be greater than the size of
universe of corresponding era, and $t_0$ is the present age of universe $\sim 10^{17}$ sec.
At the neutrino decoupling age of universe when $t\sim 1$ sec, $y\le 10^{20}$ cm,
$B_0\ge 10^{-8}$ eV. Therefore, if $B_0\sim 5\times 10^{-2}$ eV, then from
Eqs. (\ref{ospgen}) and (\ref{ospbh}) the oscillation probability is given by
\begin{eqnarray}
{\cal P}_{fEU_d}\le 0.999\,\sin^2\left(\frac{7\times10^{-4}~{\rm eV^2\,sec}}{4E}\right).
\label{bobiansp}
\end{eqnarray}
On the other hand, at GUT scale when $t\sim 10^{-35}$ sec, $y\le 10^{2}$ cm,
$B_0\ge 10^{45}$ eV $>>m_e,m_\mu,m_{e\mu}$. Therefore, the oscillation probability with the
minimum possible $B_0$
\begin{eqnarray}
{\cal P}_{fEU_{GUT}}\le \sin^2\left(\frac{1.4\times10^{8}~{\rm eV^2\,sec}}{4E}\right)
\label{bobianspgut}
\end{eqnarray}
is entirely controlled by gravitational field. Either of Eqs. (\ref{bobiansp}) and
(\ref{bobianspgut}) clearly proves that flavor oscillation is severely altered by
the gravity.

\subsubsection{Inflationary era of universe with primordial fluctuations}

We know that during inflation primordial quantum fluctuations of the spacetime is
classical and the flat Robertson-Walker expanding universe may take the form as \cite{bert,mbaryo}
\begin{widetext}
\begin{eqnarray}
ds^2= (1+2 \tilde{\phi}) dt^2 - a(t)^2 \left[\frac{\omega_i}{a(t)} dx^i dt +
\left((1+ 2\tilde{\psi}) \delta_{ij} + h_{ij}\right) dx^i dx^j \right]
\label{tens}
\end{eqnarray}
where $\tilde{\phi}$ and $\tilde{\psi}$ are scalar, $\omega_i$ are vector and $h_{ij}$
are the tensor fluctuations of the metric. Of the ten degrees of
freedom in the metric perturbations only six are independent and
the remaining four can be set to zero by suitable gauge choice.
For our application we must have atleast one nonzero $g_{ij}$, when $i\neq j=1,2,3$,
of the metric \cite{mpla} and thus
need only the tensor perturbations and we
choose the transverse-traceless (TT) gauge $h^i _i=0, \partial^i
h_{ij}=0$ for the tensor perturbations. In the TT gauge the
above perturbed Robertson-Walker metric can be expressed as
\begin{eqnarray}
\nonumber
ds^2=\left(1+2\tilde{\phi}\right)dt^2-a(t)^2\left[\frac{\omega_i}{a(t)} dx^i dt
+(1+2\tilde{\psi}-h_+)dx_1^2+(1+2\tilde{\psi}+h_+)dx_2^2\right.+\\
\left.2h_\times dx_1 dx_2 +
(1+2\tilde{\psi})dx_3^2\right].
\end{eqnarray}
\end{widetext}
Therefore, following general expression \cite{mpla,bmltst} given for any metric
the gravitational scalar potential computed for this spacetime is
\begin{eqnarray}
B^0=\partial_3 h_\times =\partial_z h_\times.
\label{bograv}
\end{eqnarray}
The gravitational scalar potential $B_0$ can be expressed as a fluctuation
amplitude $A_\times$ times a wavenumber which represents the
length scale over which the metric fluctuations vary. The Compton
wavelength of the particles in the GUT era
%($\sim (10^{16} GeV)^{-1}$)
is much smaller than the average wavelength of the gravitational waves whose wavenumber
$k \sim H = 1.66 g_*^{1/2} (T^2/M_{Pl})$ \cite{bert,mbaryo}. $H$ is the Hubble constant
at the time of horizon crossing of mode $k$. Thus gravitational wave
 background can be considered as a constant amplitude field for
 the GUT processes. Hence the mean value of $B_0$, as a function of
 temperature and the primordial tensor wave amplitude $A_\times$,
 can be expressed as
 \begin{eqnarray}
 \langle B_0 \rangle \equiv B_0 \simeq A_\times k \simeq A_\times \left(1.66~g_*^{1/2}~ {T^2 \over M_{Pl}}\right)
 \label{B0grav}
 \end{eqnarray}
Here $g_*=106.7$, is the number of relativistic degrees of freedom, for the
standard model. Primordial tensor and scalar perturbations contribute to the anisotropy of cosmic
microwave background at large angles.
The COBE DMR measurement \cite{cobe} of temperature anisotropy $\Delta T =30 \mu K$ sets an upper
limit of these fluctuation amplitudes to be $10^{-5}$. The magnitude of
the tensor perturbations depends upon the details of inflation
potential \cite{tensor} and is expected to be an
order of magnitude smaller in amplitude than scalar
perturbations. Therefore, we can set $A_\times \le 10^{-6}$. Hence, at $T\sim 10^{13}$ GeV with a
very small amplitude of fluctuation such that $A_\times\sim 10^{-19}$ the gravitational
potential comes out to be $B_0\sim 10^{-2}$ eV which is only an order of magnitude higher than
the neutrino masses. Therefore, the corresponding oscillation probability and length are given by
\begin{eqnarray}
\nonumber
{\cal P}_{fIN}&=&0.999\sin^2\left(\frac{7\times 10^{-35}\,{\rm eV^2\,sec}}{4E}\right),\\
\lambda_{fIN}&\simeq &\frac{4\pi E}{7 \times 10^{-4}\,{\rm eV^2}}
\label{osgvbsm}
\end{eqnarray}
which are an order of magnitude higher than that without gravitational effect.
At the maximum possible $A_\times$,
$B_0\sim 10^{11}\,{\rm eV}>>m_e, m_\mu, m_{e\mu}$ resulting the oscillation probability
and length
\begin{eqnarray}
\nonumber
{\cal P}_{fIN}&=&\sin^2\left(\frac{1.4\times 10^{-22}\,{\rm eV^2\,sec}}{4E}\right),\\
\lambda_{fIN}&\simeq &\frac{4\pi E}{1.4 \times 10^{-2}\,{\rm eV}\,B_0}.
\label{osgrav}
\end{eqnarray}

\subsection{Implications }

The flavor mixing and hence oscillation is influenced by gravity.
This may alter the relative abundance of different neutrino flavors in universe.
Around a primordial black hole the maximum oscillation length of
a thermal neutrino, from Eqs. (\ref{oslbh}) and (\ref{oslbhh}), is obtained as $0.54$ cm
compared to $4.6$ cm obtained from Eq. (\ref{oslf}) without gravity.

The oscillation probability increases as well significantly in early universe
when we consider gravitational effect, as understood from section IV.D. At the neutrino
decoupling era, while the probability increases for thermal neutrinos only $1.5$ times,
for TeV neutrinos it increases about two order of magnitude. It is also apparent
from Eqs. (\ref{ospf}), (\ref{bobiansp}) and (\ref{bobianspgut}) that at GUT era oscillation
takes place vigorously. Therefore, production of muon neutrinos in early universe due to oscillation
is expected to be much higher than that estimated without gravitational effect.

At the inflationary era with primordial fluctuations, the oscillation length
of, e.g., GeV neutrinos could vary from $10^8$ cm to $10^{-5}$ cm depending on $B_0$, while
size of universe is $10^3$ cm. Therefore, the oscillation would be feasible.
However, lifetime of the era itself is very small which may hinder significant oscillation.

\section{\label{dbeta}Neutrinoless double beta decay}

~~~~It is generally believed that if neutrino is its own
antiparticle, {\it i.e.} neutrino is a Majorana particle, then
neutrinoless double beta decay may be observed. As CPT violating
nature alters the mixing angle and masses in neutrino and
antineutrino sector, it is interesting to see the effect of
gravity on neutrinoless double beta decay rate. If one considers
only one flavor, then gravity does not alter the decay rate. It was
already shown \cite{bbbk} that CPT violating term has no effect
on decay rate with only one flavor. Here we investigate the case
for two flavor mixing.

If we consider neutrino flavor mixing, without
neutrino-antineutrino mixing, then one can express neutrino of
different flavors as superposition of different mass eigenstates
like (considering only electron and muon neutrino)

\begin{equation} \left( \begin{array}{c}
                        \psi_e \\
                        \psi_\mu
                       \end{array} \right)~=
~U\left( \begin{array}{c}
         f_1 \\
         f_2
         \end{array} \right)
\label{flvag} 
\end{equation} 
where $U={\cal F}(B^0=0)$, and $f_1$
and $f_2$ are mass eigenstates with masses $l_1$ and $l_2$
respectively. In this case, the amplitude for neutrinoless double
beta decay involving only electrons is \cite{doi2}
\begin{equation} A~\propto~\sum_il_iU_{ei}^2,~~~~~~~~~i=1,2.
\end{equation}
In presence of gravity the state $\Psi$ is replaced by $\nu$ as discussed
in previous sections, and so the relation given by the Eq. (\ref{flvag}) is 
modified to  
\begin{equation} \left( \begin{array}{c}
                        \nu_e \\
                        \nu_\mu
                       \end{array} \right)~=
~{\cal F}\left( \begin{array}{c}
         f_1 \\
         f_2
         \end{array} \right) 
\end{equation} 
Hence, the neutrinoless double beta decay amplitude is given by 
\begin{equation}
A_{(1,2)}\,\propto\,M_{(1,3)}\cos^2\phi_{(1,2)}\,+\,M_{(2,4)}\sin^2\phi_{(1,2)}.
\end{equation} 
From the expressions of $M$-s and $\phi$-s from Eqs. (\ref{mas4}) and 
(\ref{tana}) and using Eqs. (\ref{yy}) and (\ref{mdef}) we obtain 
\begin{equation}
A\propto\,\sqrt{B_0^2+m_e^2}.
\end{equation}
Hence, under the effect of gravity the neutrinoless double beta decay amplitude
is modified as the mass of the flavor state is modified. In absence of gravity 
the amplitude is simply proportional to $m_e$.

The double beta decay is involved with weak interactions, conventionally it 
has nothing to do gravity. Therefore, when
we compute the decay rate with $\Psi$-s, which are the states produced by 
pure charge-current interaction, it remains unchanged in presence of gravity. 
However, if we
assume that under strong gravity $\Psi$-s modify to $\nu$-s and recalculate
the rate, then it modifies by means of $B_0$.

\section{\label{sum} Summary and Discussion }

The idea of neutrino flavor oscillation is
well established both theoretically and experimentally. However,
there are some anomalies in different types of flavor oscillation
data. For example, the LSND data provides mixing angle and
neutrino masses which are inconsistent with solar and atmospheric
neutrino oscillation data. This anomaly can be removed by
introducing fourth family of sterile neutrino or CPT violation in
neutrino sector.
%In this paper, we have shown the consequences of
%CPT violation in neutrino sector.
%We have employed gravity to
%provide CPT violation which has been discussed in many earlier
%papers (e.g. \cite{mpla,dmd,bmltst}).
In a very past work,
neutrino-antineutrino oscillation was discussed in analogy
with $K^0-\bar K^0$ oscillation with maximal mixing \cite{ponte2}.
In the present paper, we have shown that in presence of gravity
neutrino-antineutrino oscillation does occur.
Neutrino and antineutrino interact differently with gravity through CPT violation.
Under this condition if the Lagrangian has a lepton number violating Majorana mass,
then neutrino-antineutrino mixing and oscillation occur.
%Mixing ratio and the mass eigenvalues
%depend on the gravitational field strength.
Without gravity, Majorana mass mixes neutrino and antineutrino states in equal ratio.
%Gravity mixes them in an unequal ratio.
Even if there is no Majorana mass, gravity splits neutrino and antineutrino in two different mass
eigenstates.
%This is because gravitational effect violates CPT.
The gravitational scalar potential $B_0$ modifies the mass of the neutrino states.
This $B_0$ behaves like a lepton number conserving mass.
The mass eigenstates, produced as a result of mixing under gravity, act as modified neutrino states.
With these states we have studied two flavor mixing and corresponding oscillation generating
four mass eigenstates.

It is quite possible that the neutrino
number what we see today is carried out off the neutrino decoupling era and the
oscillation probability, strongly influenced by gravity, at that era should determine the
relative abundance of today's flavors. We, in fact, have shown that the probability of
conversion of electron neutrino to muon neutrino has greatly been enhanced, upto a few
order of magnitude, with inclusion of gravity effect. 
%We argue that this huge
%production of muon neutrino in early universe may contribute to dark energy,
%as muon neutrino is heavier than electron neutrino.
%The present day neutrino energy can be
%interpreted as dark energy of the universe.
On the other hand, even in present age
the relative abundance of muon neutrino may be increased around primordial
black holes where gravitational field is not negligible. 
%which could be the source of dark energy as well.

It has been suggested that during core-collapse of a massive star, part of the
infalling material goes into orbit around the
compact core to form a hot, dense, centrifugally supported accretion disk
whose evolution is strongly influenced by neutrino interactions.
Under a wide range of conditions, this neutrino-dominated accretion flow
will help to produce a successful supernova explosion (see, e.g. \cite{nkp})
It will be interesting to study the effect of gravity on oscillation
in such systems and then on related supernova explosion. 
 
In early, it was shown by pure quantum field theoretical 
consideration of neutrino flavor mixing that in infinite 
volume limit, the vacuum expectation value is not invariant under the 
transformation of flavor mixing (e.g. \cite{bla,hann}). This is the 
consequence of unitary inequivalance of the flavor and mass vacua.  
It has been shown that the vacuum structure with neutrino mixing has 
non-zero contribution to vacuum energy \cite{blasone}.  
It is well known
that one of the interpretations of cosmological constant is linked with
the density ($\rho$) and pressure ($p$) of vacuum with equation of state
$p=-\rho$. The non-zero value of cosmological constant and hence the
presence of such a vacuum energy density, which is often termed as dark
energy, is needed to explain the observed acceleration of present Universe.
Thus, the non-zero flavor vacuum energy density can be interpreted as a
contribution to cosmological constant and hence to dark energy \cite{bla}.
It has been shown \cite{bla} that
the vacuum energy, the flavor vacuum expectation value of energy momentum 
tensor element $T_{00}$, is non-zero and proportional to square of one of the 
Bogoliubov coefficients $V_{\boldmath k}$ associated with the flavor 
creation and annihilation operator. In ultrarelativistic limit 
$|V_{\boldmath k}|^2$ is proportional to $(m_1-m_2)^2$, the square of the 
difference of mass eigenvalues.  Moreover, this vacuum 
expectation value is proportional to $\sin^2\phi$, where $\phi$ is the mixing 
angle.  Hence this contribution depends on the specific nature of 
the mixing. It is to be noted that in the limit 
$V_{\boldmath k}\rightarrow 0$, which is the case of traditional
phenomenological mixing, the vacuum energy vanishes. All these calculations
are in flat space. If we consider any curved spacetime, 
then the vacuum expectation
value changes accordingly. Hence the contribution to vacuum density and
then to dark energy depends on specific nature of mixing and
background metric. In the present work, we have shown that background 
curvature i.e the gravitational field affects the mixing. Hence, in this case,
we expect that gravity will play in both the ways - by affecting the mixing and 
by giving a curved 
background, to affect the contribution of dark energy. Therefore, 
it will be good to see how the present results can account for observed 
proportion of 
dark energy under the pure quantum field theoretic consideration.

With this new mixing scheme, neutrinoless double beta decay rate
has been revisited. With one neutrino flavor, it was shown
\cite{bbbk} earlier that CPT violating term has no effect on
decay rate. In this spirit we have calculated the amplitude of
neutrinoless double beta decay considering neutrino-antineutrino
admixture with two neutrino flavor mixing and still have found
that the decay amplitude is proportional to the mass of the flavor state.
However, as the mass of the flavor state, which is considered to be modified
due to gravity, is now different from that in absence of 
gravity, the decay amplitude also differs from that in absence of gravity.
Hence, the CPT violating gravity has an effect on the decay rate. 

\begin{acknowledgments}
The authors are grateful to Palash B. Pal of SINP for illuminating
discussion throughout the course of the work. They are also
thankful to Sudhir K. Vempati of IISc for discussion while writing
the paper. The authors would also like to thank the referee 
for his/her illuminating suggestions and encouragement.
\end{acknowledgments}

%\appendix
%
%\section{\label{App:anti}Particle-antiparticle mixing parameters}
%
%\begin{equation} {\cal U}_{e,\mu}~=~\left(
%   \begin{array}{cc}
%    \cos \theta_{e,\mu} & -\sin \theta_{e,\mu} \\
%
%    \sin \theta_{e,\mu} & \cos \theta_{e,\mu} \\ \end{array}
%   \right),
%    \end{equation}
%\begin{equation} \tan \theta_{e,\mu}~=~\frac{m_{e, \mu}}{B_0+\sqrt{B_0^2+m_{e,
%\mu}^2}}, \end{equation} where $m_e$ and $m_\mu$ are the Majorana
%masses for electron and muon neutrino respectively, analogous to
%$m$ of section \ref{mmatrix}.
%
%\begin{eqnarray} m_{(e,\mu)1}~=~-\sqrt{B_0^2+m_{e,
%\mu}^2}, \nonumber \\ m_{(e,\mu)2}~=~\sqrt{B_0^2+m_{e,
%\mu}^2}.\end{eqnarray}

%\section{\label{App:flavor}Flavor mixing parameters}

%\section{\label{App:T} The matrix $T$}
%\begin{widetext}
%\begin{equation} T~=~\left(
%   \begin{array}{cccc}
%    \cos \theta_e \cos \phi_1 & -\cos \theta_e \sin \phi_1 & -\sin \theta_e
%    \cos \phi_2 & \sin \theta_e \sin \phi_2\\
%
%    \cos \theta_\mu \sin \phi_1 & \cos \theta_\mu \cos \phi_1 & -\sin \theta_\mu
%    \sin \phi_2 & -\sin \theta_\mu \cos \phi_2  \\
%
%    \sin \theta_e \cos \phi_1 & -\sin \theta_e \sin \phi_1 & \cos \theta_e
%    \cos \phi_2 & -\cos \theta_e \sin \phi_2\\
%
%    \sin \theta_\mu \sin \phi_1 & \sin \theta_\mu \cos \phi_1 & \cos \theta_\mu
%    \sin \phi_2 & \cos \theta_\mu \cos \phi_2 \\ \end{array}
%   \right).
%    \end{equation} \end{widetext}
%
%\bibliography{dbeta}

%
% ****** End of file apssamp.tex ******
\end{document}